\documentclass[aps,prd,reprint,superscriptaddress,nofootinbib]{revtex4-2}
\usepackage[english]{babel}
\usepackage[utf8]{inputenc}
\usepackage[colorinlistoftodos, color=green!40, prependcaption]{todonotes}
\usepackage{orcidlink}
\usepackage{amsthm}
\usepackage{mathtools}
\usepackage{physics}
\usepackage{xcolor}
\usepackage{graphicx}
\usepackage{adjustbox}
\usepackage{placeins}
\usepackage[T1]{fontenc}
\usepackage{lipsum}
\usepackage{csquotes}
\usepackage{hyperref}
\usepackage{lipsum}
\usepackage{footnote}

\newcommand{\GC}{\textcolor{black}}

\begin{document}
\title{Neutron Stars as a Probe of Cosmic Neutrino Background}
\author{Garv Chauhan\,\orcidlink{0000-0002-8129-8034}}
    \email{gchauhan@vt.edu}
    \affiliation{Center for Neutrino Physics, Department of Physics, Virginia Tech, Blacksburg, VA 24061, USA}

\begin{abstract}
The Cosmic Neutrino Background (C$\nu$B) constitutes the last observable prediction of the standard cosmological model, which has yet to be detected directly. In this work, we show how the coherent scattering of neutrinos off dense neutron matter can lead to an additional cooling channel in neutron stars (NSs). We also include the effects of gravitational capture and boosting, but find that the cooling is efficient only in the presence of large overdensities. We further discuss the prediction of a boosted C$\nu$B flux on Earth from nearby NSs and the potential detection prospects in the case of a future nearby galactic supernova. Although currently these ideas do not offer any detection prospects, they can be used to constrain overdensities \GC{$\eta \lesssim 10^{11}\textrm{-}10^{14}$} on short length scales $\mathcal{O}(10\text{ km})$. We also discuss the impact of new physics scenarios, such as long-range forces, on NS cooling through the C$\nu$B.
\end{abstract}

\maketitle
The Cosmic Neutrino Background (C$\nu$B) is composed of relic neutrinos that decoupled from the standard model plasma at about $\sim 0.1$ seconds after the Big-Bang. Like its photon counterpart, the Cosmic Microwave Background (CMB), the C$\nu$B is also expected to encode the tiny anisotropies in the matter distribution from one of the earliest epochs in the universe. At the time of decoupling, the temperature of C$\nu$B was around $\sim 2$ MeV. Due to the expansion of the universe, the C$\nu$B has cooled down to an effective temperature of about $T_0 \sim 0.168$ meV (average momentum $|p_\nu|\simeq 3.15 T_0 $). Today, the average number density for C$\nu$B neutrino (or anti-neutrino) species per helicity mode is about $n_{\nu,\bar{\nu}} \sim 56 \text{ cm}^{-3}$. However, the extremely low average energy of C$\nu$B neutrinos makes their direct detection extremely challenging, rendering the C$\nu$B the last observable prediction of the standard cosmological model that has yet to be detected directly.

One of the earliest ideas to detect C$\nu$B is based on neutrino capture on Tritium i.e., $\prescript{3}{}{\textbf{H}}+\nu_e\rightarrow \prescript{3}{}{\textbf{He}}^{+} + e^-$, proposed by Weinberg in 1962~\cite{Weinberg:1962zza}. This idea is currently being developed by the PTOLEMY experiment~\cite{PTOLEMY:2018jst,Betti:2018bjv,PTOLEMY:2019hkd}. Various other novel detection proposals have been put forward in the literature, focusing on the Stodolsky effect~\cite{Stodolsky:1974aq,Duda:2001hd}, atomic de-excitation~\cite{Yoshimura:2014hfa}, Z-resonance~\cite{Eberle:2004ua}, accelerator~\cite{Bauer:2021uyj}, coherent scattering\cite{Opher:1974drq,Langacker:1982ih,Shvartsman:1982sn,Shergold:2021evs},  torsion balance and laser interferometry\cite{Duda:2001hd,Domcke:2017aqj}, boosting from cosmic-ray scattering~\cite{Hara:1980,Hara:1980mz,Ciscar-Monsalvatje:2024tvm,DeMarchi:2024zer,Herrera:2024upj} (also see review~\cite{Bauer:2022lri}). In this work, we will specifically look at the effects of macroscopic coherence in low-energy neutrino scattering.

The coherent neutrino-nucleon scattering was first proposed by Freedman in 1973 to probe weak neutral currents~\cite{Freedman:1973yd}. In his landmark work, Freedman had suggested that the coherence effects would be important for neutrino interactions in a neutron star (NS). This was subsequently confirmed by Wilson~\cite{Wilson:1974zz}. The effect of the coherence enhancement on the momentum transfer to a fixed target detector was calculated in Ref.~\cite{Opher:1974drq,Lewis:1979mu}, but later showed to be not viable~\cite{Cabibbo:1982bb}. In this work, we will also focus on the coherent scattering of C$\nu$B neutrinos but in the context of NSs. The effect of large coherent enhancement for C$\nu$B scattering on NS cooling was first discussed by Dixit and Lodenquai~\cite{Dixit:1983aj}. We will discuss about their calculation and aim to expand upon their work.

Our paper is organized as follows. In Section~\ref{sec:csns}, we describe the basic features of the cross-section and the coherent enhancement for neutrino-nucleon scattering in NSs. We also discuss the amount of energy exchanged during this process. In Section~\ref{sec:theory}, we outline the theoretical setup for the proper calculation of coherent neutrino-nucleon scattering inside NSs. We motivate the calculation of desired dynamic structure factors for low-energy C$\nu$B neutrinos in ultra dense nuclear matter. In Section~\ref{sec:boost}, we examine three different cases for production of energetic C$\nu$B flux due to scattering off with NS and discuss the resultant neutrino flux at Earth. In Section~\ref{sec:longrange}, we investigate the effect of new physics scenarios, such as long-range forces, on the NS cooling through C$\nu$B. Finally, we conclude and outline future directions in Section~\ref{sec:conclusion}.

\section{Coherent Scattering and Neutron Stars}\label{sec:csns}
Coherent scattering usually occurs when the de Broglie wavelength of the incoming particle is larger than the typical size of the target $R_T$ and the magnitude of momentum transfer $\textbf{q}$ is small i.e. $|\textbf{q}|R_T \lesssim 1$~\cite{Akhmedov:2018wlf,Shergold:2021evs}. For an incoming neutrino with a macroscopic wavelength in a usual fixed target scattering experiment, the coherent cross-section per nucleon scales with target density $\rho$. Therefore, the total cross-section scales as $\rho^2$~\cite{COHERENT:2017ipa}. For terrestrial experiments, the densest possible elements include Lead ($\prescript{207}{82}{\textbf{Pb}}$) and Osmium ($\prescript{190}{76}{\textbf{Os}}$). While the enhancement in coherent neutrino-nucleon cross-section from $(A-Z)^2$ i.e. number of neutrons is higher for Lead, the density enhancement is larger for osmium by almost a factor of 4. However, in spite of coherent enhancement, the small momentum transfers limit the effect of C$\nu$B on the target material. 

So we ask ourselves, can we enhance the scattering rate even further or enhance the effect of C$\nu$B neutrinos on the target it scatters off of? As we have seen, the maximum terrestrial elemental densities can reach only up to $\sim 22 \text{ g cm}^{-3}$ (Osmium ($\prescript{190}{76}{\textbf{Os}}$)). However, NSs are far denser than any target material available on Earth. The typical matter density in a NS can reach $\sim 10^{14} \text{ g cm}^{-3}$ and are known to be the densest objects in the universe. Naively this should lead to an overall enhancement of 26 orders of magnitude for the total cross-section.

Let us first try to estimate the cross-section for the C$\nu$B interacting with the dense nuclear medium inside the NSs. The coherent neutrino-nucleus cross-section for a given target nucleus is given by \cite{COHERENT:2017ipa,Shergold:2021evs,Bednyakov:2018mjd}
\begin{equation}
    \sigma_{\nu\text{-}N} = \frac{G_F^2}{4\pi}(A-Z)^2 E_\nu^2
    \label{eq:sigma0}
\end{equation}
while the enhancement factor from macroscopic coherence (i.e. number of neutrons inside the coherent volume) will be 
\begin{equation}
    \eta_{\text{coh}}= \left(\frac{2\pi}{p_\nu}\right)^3 \frac{\rho_\text{NS}}{m_\text{N}}
\label{eq:cohfac}
\end{equation}
assuming momentum transfers of order incoming neutrino momentum, $|\textbf{q}|\sim p_\nu$. \GC{For our discussion, we will assume a momentum distribution for C$\nu$B neutrinos peaked at $p_\nu\simeq 3.15\,T_0$, where $T_0 \sim 0.168$ meV. This is an excellent approximation for non-relativistic massive neutrinos~\cite{Vitagliano:2019yzm}}\footnote{Although after the C$\nu$B turns non-relativistic, the cosmological expansion does not preserves the Fermi-Dirac distribution. However for our case, it suffices to treat the C$\nu$B distribution as Fermi-Dirac distribution with effective temperature $T_0$~\cite{Trautner:2016ias}.}. 
Therefore, for typical values of the parameters assuming $\rho_\text{NS} \simeq 10^{14} \text{ g cm}^{-3}$
and setting $m_\nu = 50$ meV. We also assume the NS medium is composed solely of neutrons i.e. $A=1, Z=0$, extending to the case of other species is straightforward. \GC{Therefore, we obtain following values for Eq.~\ref{eq:sigma0} and ~\ref{eq:cohfac}}
\GC{\begin{equation}
    \sigma_{\nu\text{-}N} \simeq 10^{-59} \text{ cm}^{2},\quad \eta_{\text{coh}} \simeq 10^{35}
\end{equation}}
Therefore, the total coherent elastic cross-section per neutron comes out to be 
\begin{equation}
    \sigma_{\nu\text{-}\text{N}} \simeq 10^{-24} \text{ cm}^{2}
    \label{eq:crossnum}
\end{equation}
This cross-section far exceeds the geometric cross-section of a neutron i.e. $\text{R}_\text{N}^2 = (0.8\text{ fm})^2 \simeq 10^{-26} \text{ cm}^{2}$. Such a feature is only possible for long-ranged interactions like gravitation. Since the neutrino-nucleus scattering is mediated by Z boson, this implies the calculated cross-section cannot exceed the geometric cross section. This points to the fact that possibly all of the C$\nu$B will interact with the NS. Note that not only the cross-section per nucleon increases, but also the number of targets/nucleons are also extremely high that aids the total cross section. If we require at least one interaction for each C$\nu$B neutrino as it traverses through the NS, the required cross-section ($\sigma_{\nu\text{-}N}$) for the process can be estimated as follows
\begin{equation}
    N_{\text{int}} \sim n_\text{NS}\, \sigma_{\nu\text{-N}}\, L_{\text{NS}}
\end{equation}
where $n_N$ is the neutron number density inside the NS core and $L_{\text{NS}}$ is the traversed length $\simeq 10$ km. Therefore, 
\begin{align}
    \sigma_{\nu\text{-N}} \simeq \frac{1}{n_\text{N} L_{\text{NS}}} \simeq 10^{-44} \text{ cm}^2
\end{align}
On comparing this cross-section with Eq.~\ref{eq:crossnum}, we can clearly observe that the effect from naive coherent enhancement far exceeds the required minimal cross-section to scatter at least once by 20 orders of magnitude. For later discussion, we also note that the neutrino-nucleon cross-section for $m_\nu=50$ meV without any macroscopic coherence effects is nearly $10^{-59} \text{ cm}^2$.

While both the Ref.~\cite{Dixit:1983aj} and our work relies on coherent enhancement, the former does not discuss details about the cross-section itself. Even our quantitative discussion above lacks to account for various effects such as Pauli blocking of final state neutron and strong nucleon-nucleon correlations in the NS. In this work, we will also shed light on these details and discuss the steps to pave way for a proper calculation of this process in the NS medium.

As also first pointed out in Ref.~\cite{Dixit:1983aj}, at current epoch C$\nu$B effectively constitutes an extremely cold Fermi gas, while the NS is a hot thermal medium. As argued earlier due to the large expected cross-section for C$\nu$B interaction with the NS, the C$\nu$B neutrinos will eventually exchange energy with the NS medium through scatterings. Hence, C$\nu$B interacting with a NS will lead to NS cooling. 

Neutrons form a highly degenerate Fermi gas in the interior of a NS. Naively based on the physics of degenerate Fermi gas at zero temperature, it might be expected that no energy exchange should occur due to the severe Pauli blocking of the final state neutron. However, the non-zero temperature environment plays an important role here. 
\begin{figure}[t]
    \centering
    \includegraphics[width=\columnwidth]{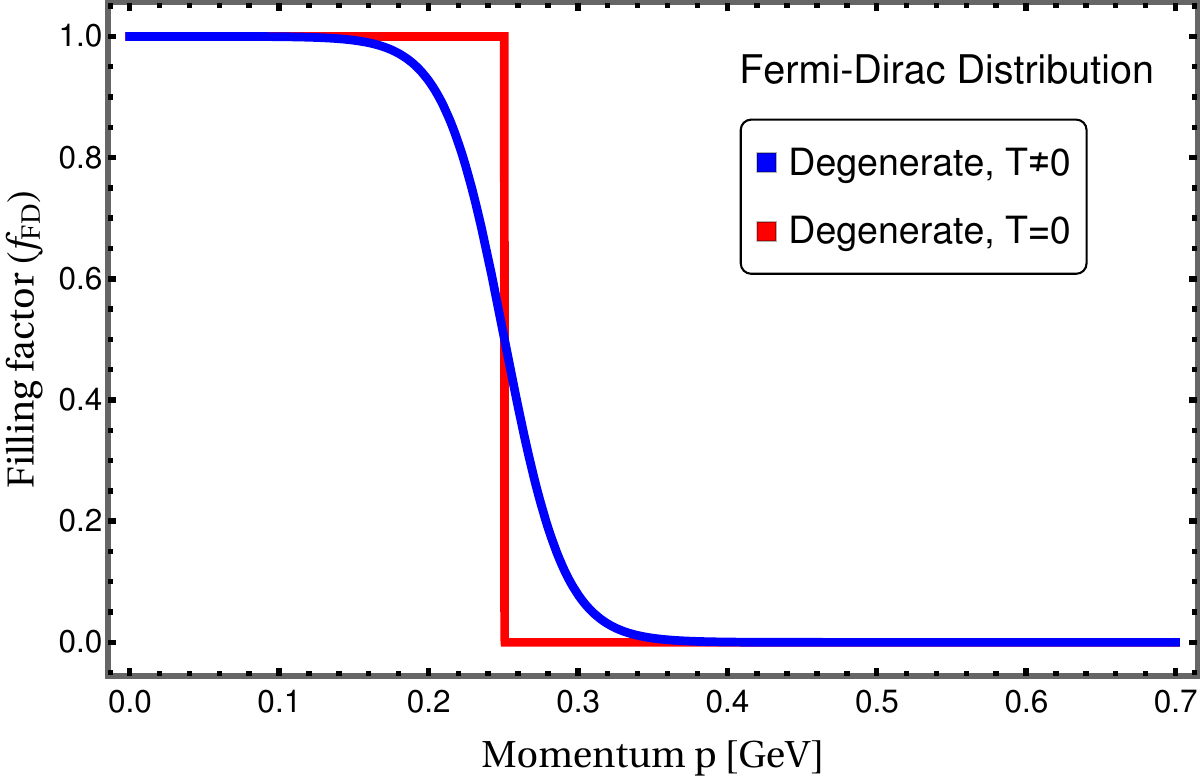}
    \caption{Fermi-Dirac distribution for degenerate neutron gas at $T=0$ and $T\neq0$.}
    \label{fig:fd}
\end{figure}
In fig.~\ref{fig:fd}, the red curve shows the Fermi-Dirac distribution for degenerate Fermi gas at $T=0$ (assuming chemical potential $\mu_N=250$ MeV, excluding rest mass), where all the momentum states upto the Fermi sphere are occupied. However for the blue curve at $T\neq0$, there is a finite number of available states to be filled near the Fermi sphere. For e.g. a neutron from state near $p=\mu_N$ can scatter into a lower momentum state with energy exchange of $\mathcal{O}(T)$ that leads to only a modest suppression factor $\sim 0.27$ i.e. $\left(1-\frac{1}{e^{-1}+1} \right)$. Therefore, on average scattered C$\nu$B neutrinos can gain energy $\mathcal{O}(T)$. This further adds to justification for the assumption made in Ref.~\cite{Dixit:1983aj}. 

Therefore, an appropriate estimate of the coherence factor is obtained by taking into account only those momentum states which can easily scatter. This corresponds to a region of finite thickness $\mathcal{O}(T)$ at the top of the Fermi sphere. While this is a good assumption for large momentum transfers $\sim T$, the coherence factor for small momentum transfers will likely be underestimated but will be used to obtain conservative estimate for the coherence factor.

For a complete degenerate gas with chemical potential $\mu$ and particle mass $m$, integrating over all momentum states upto $\sqrt{\mu^2-m^2}$ yields the total particle number density. For estimating realistic coherence factor, we will integrate 
\begin{equation}
    \eta_{\text{coh}}= \left(\frac{2\pi}{|\textbf{q}|}\right)^3 \frac{g}{(2\pi)^3} \int_{\sqrt{\mu^2-m^2}-T}^{\sqrt{\mu^2-m^2}+T} dp\frac{4\pi p^2}{e^{\frac{\sqrt{p^2+m^2}-\mu}{T}}+1}
\end{equation}
Note that we recover Eq.~\ref{eq:cohfac} if the lower and upper integration limits are restored to $0$ and $\infty$ respectively. For understanding the above result numerically, we choose two different temperatures of $100$ keV and $1$ keV  with fixed relativistic chemical potential $\mu_N = m_N + 30$ MeV. The $\mu_N$ chosen here reproduces $\rho_N \simeq 10^{14} \text{ g cm}^{-3}$. The representative temperatures chosen in these two cases will be useful for Section~\ref{sec:boost}. We will first solely focus on the case of low-momentum exchange in the first few scatterings i.e. $|\textbf{q}|\sim p_\nu$. For $T=1$ keV and $100$ keV, we find the $\eta_{\text{coh}} \sim 10^{31}$ and $10^{33}$ respectively. These coherence factor should be compared with the naive enhancement factor calculated using Eq.~\ref{eq:cohfac}, $\eta_{\text{coh}} \simeq 10^{36}$. We conclude that despite the effect of strong Pauli blocking (decreasing $T$ while holding $\mu_N$ fixed), the coherence factor does not decrease drastically. Even for $T=1$ keV case, the $\eta_{\text{coh}}$ is far higher than what is required for single scattering scenario.

In the case of coherent scattering leading to exchange of only small momentum transfer  with the NS medium in each scattering, the increasing energy of the neutrino increases the cross-section as $E_\nu^2$ while the coherence factor goes down as $1/p_\nu^3$ with each successive scatter. It can be easily estimated that coherent scattering rate even with momentum transfers up to a fraction of temperature $T$ are quite high in the degenerate neutron medium.
\begin{align}
    \sigma_{\nu-N} & = \frac{G_F^2}{4\pi}E_\nu^2 \frac{g}{|\textbf{q}|^3} \int_{\sqrt{\mu^2-m^2}-T}^{\sqrt{\mu^2-m^2}+T} dp\frac{4\pi p^2}{e^{\frac{\sqrt{p^2+m^2}-\mu}{T}}+1} \nonumber \\
    & = \frac{G_F^2}{2\pi}\frac{1}{f T} \int_{\sqrt{\mu^2-m^2}-T}^{\sqrt{\mu^2-m^2}+T} dp\frac{4\pi p^2}{e^{\frac{\sqrt{p^2+m^2}-\mu}{T}}+1}    
\end{align}
where we have used $E_\nu=\sqrt{m_\nu^2+|\textbf{q}|^2},\, |\textbf{q}| = f T$, where $f \in [p_0/T,1]$ and yields the cross-sectional area  
\begin{equation}
     \sigma_{\nu-N} \simeq 5 \times 10^{-41} \text{ cm}^2 f^{-1} ,\, \text{(for both $T=1$ and $100$ keV)}
\end{equation}
which at minimum is at least 3 orders of magnitude higher than required for the case of a single scatter within the NS core. 

\section{Coherent Neutrino-Nucleon Scattering in NS}\label{sec:theory}
We will discuss some more details about the mechanism that might lead to the \textit{thermalization} of C$\nu$B\footnote{Using \textit{thermalization} might be abuse of terminology, since C$\nu$B likely would not be trapped to thermalize completely. However, we will use this term to avoid being verbose.}. Two major ingredients are mainly responsible for the thermalization : $(1)$ Effective temperature of C$\nu$B ($1.95$ K) being colder than the core temperatures in NSs ($10^{6}-10^{8}$ K) $(2)$ High neutron densities coupled with large de Broglie wavelength of the C$\nu$B leads to an extremely large coherence factor for the scattering cross-section. 

In a medium, the general expression for cross-section per nucleon for neutral current scattering of low-energy neutrinos on a gas of non-relativistic neutrons or protons is given by~\cite{Sawyer:1975js,Iwamoto:1982zp,Reddy:1998hb,Horowitz:2006pj,Horowitz:2016gul,Bedaque:2018wns,Schuetrumpf:2020a}
\begin{widetext}
\begin{equation}
    \frac{1}{N}\frac{d^2\sigma_{\text{total}}(E_\nu)}{d\cos\theta\,dq_0}= \frac{G_F^2}{4\pi^2}\,(E_\nu-q_0)^2\,(C_V^2(1+\cos{\theta})S_V(\textbf{q},q_0)\,+ \, C_A^2(3-\cos{\theta})S_A(\textbf{q},q_0))
    \label{eq:genScattering}
\end{equation}
\end{widetext}
where $G_F$ is the Fermi coupling constant, $N$ is the number of target neutrons/proton, $E_\nu$ is the incoming neutrino energy, $|\textbf{q}|$ is the magnitude of momentum transfer, $q_0$ is the amount of energy transfer to the medium (negative value indicates energy gain by the neutrino), $C_V$ and $C_A$ are the vector and axial vector coupling constants and $S_{V,A}(|\textbf{q}|,q_0)$ are the dynamic structure factors. Note that the momentum transfer from kinematics is constrained to be $|\textbf{q}|=\sqrt{4 E_\nu(E_\nu-q_0)\sin^2(\theta/2)+q_0^2}$. 

In general, the dynamic structure factors depends on the complex nuclear structure inside the NS cores. The general expressions for the dynamic structure factors are 
\begin{align}
S_V(\textbf{q},q_0) &= \frac{1}{2 \pi n} \int dt\, e^{iq_0t}\langle \Phi_0|\hat{\rho}(\textbf{q},t)\hat{\rho}(-\textbf{q},t)|\Phi_0\rangle \\
S_A(\textbf{q},q_0) &= \frac{2}{3 \pi n} \int dt\, e^{iq_0t}\langle \Phi_0|\hat{s}(\textbf{q},t)\hat{s}(-\textbf{q},t)|\Phi_0\rangle \\
\hat{\rho}(\textbf{q},t) &= \frac{1}{V}\sum_{i=1}^{N}e^{i\textbf{q}\textbf{r}_i}, \quad \hat{s}(\textbf{q},t) = \frac{1}{V}\sum_{i=1}^{N} \hat{s}_ie^{i\textbf{q}\textbf{r}_i} 
\end{align}

where $\Phi_0$ is the ground state wave-function for the coherent volume. The latter two expressions essentially provide the Fourier transforms of density and spin-density operators. The usual computation of the dynamic structure factors either require a fully time-dependent simulation or capability to resolve the full excitation spectrum of the system~\cite{Schuetrumpf:2020a}. It is generally seen that the axial dynamic structure factor drops a lot due to high densities leading to enhanced ion screening from electrons. 
Therefore, we will just focus on vector type coupling.

For the terrestrial case of non-interacting nucleons and usual assumption of elastic scattering, the cross-section can be defined in terms of static structure factors. Firstly, we see that the static structure factors are defined in terms of dynamic structure factors while integrating over all kinematically allowed energy transfers, 
\begin{align}
    \int_{-|\textbf{q}|}^{\text{min}(2E_\nu-|\textbf{q}|,|\textbf{q}|)} dq_0\, S_{V,A}(\textbf{q},q_0)&\simeq\int_{-\infty}^{\infty} dq_0\, S_{V,A}(\textbf{q},q_0) \nonumber\\ &=S_{V,A}(\textbf{q})
    \label{eq:Sstatic}
\end{align}
where the first equality holds only if significant part of the response lies in the kinematically allowed region for $q_0$. Usually for the case of thermal neutrons, where $m_N\gg T$, the static structure factors are a good approximation. In this other way, we can write the differential cross section for say the vector contribution as follows 
\begin{widetext}
\begin{equation}
    \frac{1}{N}\frac{d\sigma(E_\nu)}{d\cos\theta}= \frac{G_F^2}{4\pi^2}\,(E_\nu)^2\,(C_V^2(1+\cos{\theta})S_V(\textbf{q}))
\end{equation}
\end{widetext}
The above expression is the cross-section for single neutrino-nucleon cross section multiplied by $S(\textbf{q})$. Therefore, it is clear that the $S(\textbf{q})$ contains the contributions from coherence and indeed is proportional to the coherent structure factors used in low-energy coherent elastic scattering~\cite{Akhmedov:2018wlf,Shergold:2021evs,Bauer:2022lri}. Note that in $\textbf{q}\rightarrow\infty$ limit, $S(\textbf{q})=1$ and coherent enhancement is completely lost.


Given the relevant dynamic structure factors, two different thermalization scenarios might arise. In the first case for the incoming low-energy C$\nu$B, the structure functions might peak at smaller momentum transfers (as in the usual terrestrial case). The C$\nu$B then undergoes multiple scatterings in the NS medium to reach an average energy $\sim T$. This is the case discussed earlier in the previous section. However in the second case, it might happen that due to the high densities and the strong nucleon-nucleon interactions, the dynamic structure functions can peak at higher momentum transfers. This might directly lead to the thermalization of C$\nu$B within a few scatters. As discussed in the previous section, the drop in coherence for high $|\textbf{q}|$ is somewhat overcome by energy transfer i.e. $\propto (E_\nu+|\textbf{q}|)^2$ (see Eq.~\ref{eq:genScattering}). Therefore, it can be concluded on general grounds that most likely C$\nu$B will be thermalized with the NS core.

\section{Neutron Star Cooling}
The NSs are composed of different layers of increasing densities. The outermost layers \textit{ocean} and \textit{outer crust} consists of electrons and nuclei, each typically extending upto $\sim 100$ m. While the typical densities in the ocean can be $\mathcal{O}(10^4 \text{ g cm}^{-3})$, the densities in the outer crust can reach $\mathcal{O}(10^{11} \text{ g cm}^{-3})$ with nuclei arranged into a crystalline lattice. At a density of around $\rho_{\text{ND}}=4\times 10^{11} \text{ g cm}^{-3}$ at the base of the outer crust, the neutrons start to drip out of nuclei and form a neutron gas between the nuclei. This layer forms the \textit{inner crust} and can extend nearly upto a km and densities reaching upto $0.5\,\rho_0$. Here $\rho_0=2.8\times 10^{14} \text{ g cm}^{-3}$ is the nuclear saturation density. The \textit{outer core} has densities ranging from $0.5\,\rho_0$ - $2\,\rho_0$ and consists mostly of neutrons along with protons, electrons and muons. It ranges several kilometers and constitutes most of the NS mass. In heavier NSs at even higher densities ($>2\,\rho_0$) the composition of the \textit{inner core} might differ than the outer core. Currently, theoretical understanding of inner core at such extreme densities is still under study. 

Although we have discussed a detailed view of the NS composition, for our purposes we will model the interior of the NS as composed of an isothermal core extending upto $\sim 10$ km. The internal temperature throughout the NS will be denoted as $T_{\text{core}}$. While the internal temperature of newly born NSs $\sim \mathcal{O}$(MeV), the oldest observed NSs have cooled down to just a few keV. Although the NSs are the hottest after their birth, cooling through C$\nu$B might only be competitive for the oldest NSs. 

In absence of a complete understanding of the structure function for such low-energy neutrinos in presence of strong nucleon-nucleon correlation in the dense nuclear matter and due to a general expectation of large coherent enhancement in the cross-section, we will hereafter assume $100\%$ flux of captured C$\nu$B is thermalized~\cite{Dixit:1983aj}. As discussed earlier, initially C$\nu$B is at a cold temperature $T_0$ as compared to the NS core temperatures $T_{\text{core}}$, therefore we will also assume thermodynamically C$\nu$B will gain energy of $\mathcal{O}(T_{\text{core}})$ on average.
\begin{figure}[t]
    \centering
    \includegraphics[width=\columnwidth]{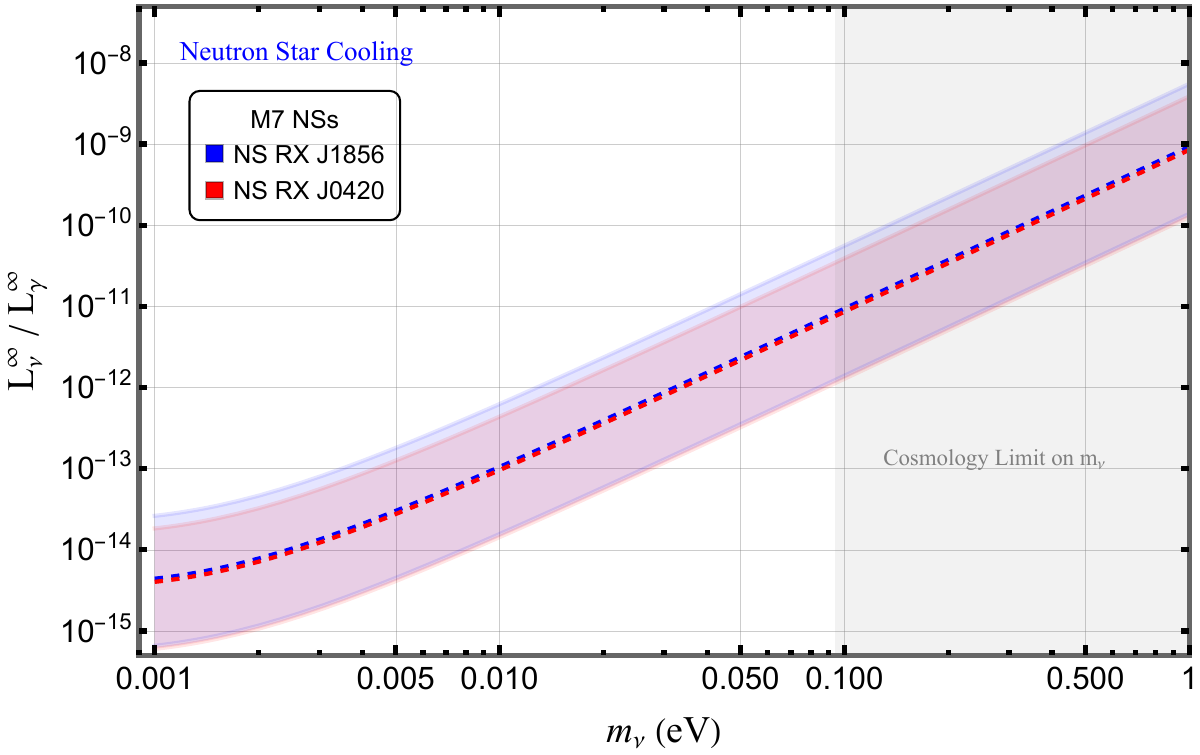}
    \caption{The ratio of energy loss through C$\nu$B ($L_{\text{C$\nu$B}}^{\infty}$) interactions over surface cooling through photons ($L_{\gamma}^{\infty}$). The ratio $L_{\text{C$\nu$B}}^{\infty}/L_{\gamma}^{\infty}$ directly constrains the overdensities of C$\nu$B near cold NSs, see text for details.}
    \label{fig:p1}
\end{figure}
\begin{table}
    \centering
    \begin{tabular}{|c|c|c|c|}
    \hline
        ID & Distance (in pc) & $T_{\text{sur}}^{\infty}$ (in eV) & $T_{\text{core}}^{\infty}$ (in keV) \\ \hline
        J1856 & $123 \pm 13$  & $50 \pm 14$  & $5 \pm 3$ \\
        J0420 & $345 \pm 200$ & $45 \pm 10$ & $3 \pm 2$ \\
    \hline
    \end{tabular}
    \caption{The distance to Earth, surface ($T_{\text{sur}}^{\infty}$) and core boundary temperatures ($T_{\text{core}}^{\infty}$) for the two Magnificient Seven NSs used in this work~\cite{Haberl:2006xe,Potekhin:2020ttj,Buschmann:2019pfp}. \GC{Note that $T_{\text{core/sur}}^{\infty}$ are defined as the core/surface temperatures for a distant observer at infinity.}}
    \label{tab:M7}
\end{table}

A NS is an extremely dense object. Therefore, C$\nu$B reaching the surface of the NS falls through a deep gravitational potential well, which increases the average kinetic energy of the infalling neutrinos. Outside the potential well, average kinetic energy $\langle E_{\nu,0}^{\text{kin}}\rangle \sim T_0$. While near the NS surface, the average kinetic energy increases to $\langle E_{\nu}^{\text{kin}}\rangle \sim \langle E_{\nu,0}^{\text{kin}}\rangle + \frac{\text{G} M_{\text{NS}}}{R_{\text{NS}}}m_\nu$. It can be seen that energy gain is suppressed by the tiny neutrino masses. Therefore, the de Broglie wavelength that determines the coherent volume will not change appreciably while the incoming neutrino will gain more velocity.
The average velocity of the incoming flux reaching the NS surface denoted here by $v_{\text{boost}}$ can be approximated by the average of C$\nu$B velocity far away from the NS and velocity at the NS surface
\begin{equation}
    v_{\text{boost}}= \frac{1}{2}\left(v_{\text{avg}} + \sqrt{v_{\text{avg}}^2+ \frac{G M_{\text{NS}}}{R_{\text{NS}}}} \right) \simeq \frac{1}{2}\sqrt{ \frac{G M_{\text{NS}}}{R_{\text{NS}}}}
\end{equation}
where $v_{\text{avg}}$ is the average velocity of the C$\nu$B neutrinos. We will use standard values for $M_{\text{NS}}=1.4 M_{\odot}$ and $R_{\text{NS}}= 10$ km throughout the rest of this work. In general relativistic terms, the Lorentz factor in the local frame near the NS surface is given by
\begin{equation}
    \gamma'_{\text{NS}}= \gamma_{0}\sqrt{ 1+ 2\frac{G M_{\text{NS}}}{R_{\text{NS}}}}
\end{equation}
We also note that in extremely old and cold NSs, if average thermal energy of neutrons is less than the infalling neutrinos, it will instead lead to heating of the NSs. Since given the current cosmological bound on neutrino masses ($\Sigma m_\nu < 0.26$ eV), heating scenario will require extremely cold NSs which have not been observed yet, so we do not consider this possibility in our work. 

Another important effect occurs due to the strong gravitational pull of the NS. In absence of gravity, the capture cross section is $ \sigma_{\text{cap}} \sim 4\pi R_{\text{NS}}^2$. As mentioned earlier the long range nature of the gravitational force can lead to an enhancement of the capture cross-section, which can be larger than the geometric cross-section. Therefore for slow moving C$\nu$B, the effective capture cross-section depends on the initial velocity distribution far away from the NS. For mono-energetic neutrinos (a good approximation for non-relativistic C$\nu$B),  the gravitational capture factor can be calculated simply using energy and momentum conservation. For a given spectrum, the capture enhancement can be calculated as in Ref.~\cite{Gould:1987ir,Goldman:1989nd}. The enhancement factor is given by
\begin{equation}
     \sigma_{\text{cap}} \sim 4\pi R_{\text{NS}}^2 \left(1+\frac{v_{\text{esc}}^2}{v_{\text{avg}}^2}\right)
\end{equation}
where $v_{\text{esc}}$ is the escape velocity at the surface of the NS. Note in our calculation, we assume the NS to be at rest with respect to the C$\nu$B frame. The calculation for large relative NS velocities will tend to the $m_\nu\rightarrow 0$ limit of the former case. 

Finally we will calculate the total energy loss assuming energy transfer $\mathcal{O}(T_{\text{core}})$ for the entire captured flux of cosmic neutrino background with number density $n_\nu$. The energy loss rate through interactions with C$\nu$B \GC{in the local NS frame} is given by
\begin{widetext}
\begin{equation}
    L_{\text{C$\nu$B}}= (4\pi R_{\text{cap}}^2)\times(n_\nu v_{\text{boost}})\times\Delta E  = 4 \pi\, R_{\text{NS}}^2   \left(1+ \frac{v_{\text{esc}}^2}{v_{\text{avg}}^2}\right)\times n_{\nu}v_{\text{boost}}\times k_B\,T_{\text{core}}
\end{equation}
\label{eq:CNBcool}
\end{widetext}
where $R_{\text{cap}}$ is the effective capture radius of the NS including the enhancement from gravitational capture. \GC{Transforming this energy loss rate from local NS frame to the frame for an observer at infinity, we obtain
\begin{widetext}
\begin{equation}
    L_{\text{C$\nu$B}}^\infty = L_{\text{C$\nu$B}}^{core}\left(1-2\frac{G M_{\text{NS}}}{R_{\text{NS}}}\right) = 4 \pi\, R_{\text{NS}}^2   \left(1+ \frac{v_{\text{esc}}^2}{v_{\text{avg}}^2}\right)\times n_{\nu}v_{\text{boost}}\times k_B\,T_{\text{core}}^{\infty}\sqrt{1-2\frac{G M_{\text{NS}}}{R_{\text{NS}}}}
\end{equation}
\label{eq:CNBcool2}
\end{widetext}}
\GC{where one factor of gravitational redshift has been absorbed into the NS core temperature at infinity i.e. $T_{\text{core}}^{\infty} = T_{\text{core}}\sqrt{1-2\frac{G M_{\text{NS}}}{R_{\text{NS}}}}$~\cite{Potekhin:2020ttj}. For our chosen values of the neutron star mass and radius, its value is $\sim 0.76$.}

In the standard scenario, a NS cools down in its initial stages through neutrino emission from the core. The neutrino emission stage can last up to $\sim 10^5$ years. The photon cooling stage begins thereafter, when the neutrino energy losses falls below the energy loss through surface thermal emission in photons~\cite{Hansen:1997hb,Yakovlev:2004iq,Yakovlev:2004yr,Potekhin:MainReview}. 

In the neutrino emission stage for standard slow cooling scenario, the main processes involved are modified Urca and $NN$-bremsstrahlung (where $N$ is a nucleon). The neutrino energy loss rate scales as $Q_{\text{mURCA}},Q_{\text{$\nu$-Brem}}\propto T^8$. Since $Q_{\text{C$\nu$B}}$ only scales as $T$, this cooling channel cannot be consequential during the neutrino cooling stage~\cite{Dixit:1983aj}. Therefore realistically $Q_{\text{C$\nu$B}}$ should be compared with the luminosity loss through surface emission through photons at temperature $T_{\text{sur}}$ ($\ll T_{\text{core}}$)
\GC{\begin{equation}
    L_{\gamma}^{\infty}= 4 \pi R_{\text{NS}}^2\,\sigma_{\text{SB}}\,(T_{\text{sur}}^{\infty})^4
\end{equation}}
where $\sigma_{\text{SB}}$ is the Stefan-Boltzmann constant~\cite{Potekhin:2020ttj}. 

For our purposes, we will focus on few of the coldest candidates in the Magnificient Seven (M7) NSs, namely RX J1856.5-3754 (J1856) and RX J0420.0-5022 (J0420)~\cite{Haberl:2006xe}. M7 is a nearby group of seven isolated NSs characterized by soft thermal X-ray emission. The details for their distance to Earth, surface and core boundary temperatures are given in Table~\ref{tab:M7}~\cite{Potekhin:2020ttj,Buschmann:2019pfp}.

In fig.~\ref{fig:p1}, we plot the ratio of total energy loss through C$\nu$B neutrinos ($L_{\text{C$\nu$B}}$) to energy loss through surface emission of photons ($L_{{\gamma}}$) as a function of neutrino mass ($m_\nu$). The dependence on $m_\nu$ arises from the indirect dependence of the gravitational capture enhancement on $m_\nu$ (through $v_{\text{avg}}$). Since the lightest neutrino mass scale is unknown, the masses for all three mass eigenstates are not fixed and we choose to plot it as a function of $m_\nu$. The blue and red bands signify the 1-$\sigma$ band for constraining the ratio $L_{\text{C$\nu$B}}/L_{\gamma}$ from J1856 and J0420 respectively. Therefore, it can be clearly seen that C$\nu$B cooling has essentially no effect on the NS, as also found in Ref.~\cite{Dixit:1983aj}. 

It should be noted that inverse of the quantity $L_{\text{C$\nu$B}}/L_{{\gamma}}$ gives a direct probe of the C$\nu$B overdensities at the NS location, denoted simply by $\eta$ (not to be confused with coherence factor, which is denoted by $\eta_{\text{coh}}$). Using the criterion that $L_{\text{C$\nu$B}}/L_{{\gamma}} \simeq 1$ can be constrained from NS cooling measurements~\cite{Dixit:1983aj}, we find that the cooling through C$\nu$B can constrain maximum overdensities around $10^{10}$ for $m_\nu \sim 0.1$ eV.
Although Pauli exclusion limits $\eta$ disfavors $\eta> 10^{6}$ but it might change in presence of non-standard scenarios~\cite{Bauer:2022lri,Smirnov:2022sfo,Batell:2024hzo}. The current model-independent terrestrial bounds from KATRIN limit $\eta < 10^{11}$~\cite{KATRIN:2022kkv}. We observe NS cooling argument sets better limits on the overdensities. However, it should be noted that overdensities around the NS need not be same as on Earth. 

Recently, the astrophysical neutrino fluxes scattering off of C$\nu$B have been used to constrain overdensities~\cite{Brdar:2022kpu}. The C$\nu$B scattering with cosmic rays and neutrino flux from TXS 0506+056 can yield a boosted relic neutrino flux, the non-observation of these events yield limits $\eta < 10^{13}$ in the Milky way and $\eta < 10^{10}$ in TXS 0506+056~\cite{Ciscar-Monsalvatje:2024tvm}. Similarly limits around $\eta < 10^{8}$ can be placed from the non-attenuation of the high-energy neutrino flux from NGC 1068~\cite{Franklin:2024enc} and weighted $\eta < 10^{10}$ from C$\nu$B scattering with cosmic ray reservoirs~\cite{DeMarchi:2024zer}. However, all these methods constrain overdensities on length scales around kpc-Gpc. We have shown that NS cooling through C$\nu$B can constrain overdensities \GC{$\eta \lesssim 10^{11}\textrm{-}10^{14}$ (dependent on $m_\nu$)} on the shortest length scales  $\mathcal{O}(10)$ km.

\section{Boosted C$\nu$B flux}\label{sec:boost}
\begin{figure*}[t]
    \centering
    \includegraphics[width=1.9\columnwidth]{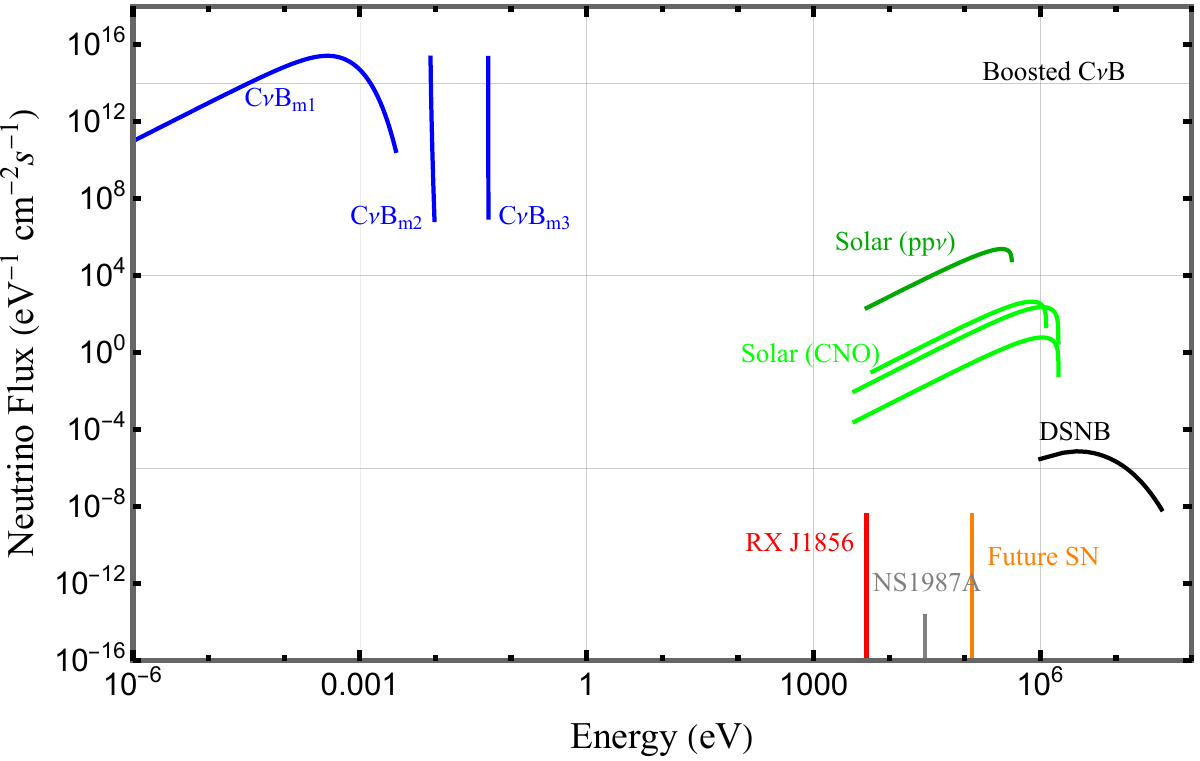}
    \caption{The boosted relic neutrino flux (in units of $\text{eV}^{-1}\text{cm}^{-2}\text{s}^{-1}$) from different sources for $m_\nu = m_3 =50$ meV : NS RXJ1856 (red), NS1987A (grey) and for a nearby future SN remnant (orange). For discussion on the temperature and width of the boosted spectrum, see text. We also plot few other relevant components of the Grand Unified Neutrino Spectrum at Earth for comparison~\cite{Vitagliano:2019yzm} : solar pp-chain (dark green), solar CNO (green), diffuse supernova neutrino background labeled DSNB (black) and standard C$\nu$B flux (blue) assuming minimal mass scenario for normal ordering of neutrino masses ($m_1=0$ meV, $m_2=8.6$ meV, $m_3=50$ meV).}
    \label{fig:p3}
\end{figure*}
As studied in the previous section, cooling a NS through C$\nu$B in the standard scenario has potentially no observable effect even in the oldest NSs. However after the thermalization, the NS boosted C$\nu$B neutrinos might lead to an observable neutrino flux at Earth. 

Using the Eq.~\ref{eq:CNBcool}, the total boosted C$\nu$B flux at Earth is given by 
\GC{\begin{equation}
     \Phi_\nu^{\text{Earth}} = n_{\nu}v_{\text{boost}}\,  \left(1+ \frac{v_{\text{esc}}^2}{v_{\text{avg}}^2}\right)\,\left(\frac{R_{\text{NS}}}{d_{\text{NS}}}\right)^2\sqrt{1-2\frac{G M_{\text{NS}}}{R_{\text{NS}}}}\,
\end{equation}}
where $d_{\text{NS}}$ is the distance of the NS to Earth. By scaling the standard low-energy C$\nu$B flux on Earth by gravitational capture factor and the average velocity, the differential boosted flux approximately can be written as 
\GC{\begin{align}
    \frac{d\Phi_\nu^{\text{Earth}}}{dE} &= \frac{1}{2\pi^2 v_{\text{avg}}}\frac{{(E-T_{\text{core}})^2-m_\nu^2}}{e^{\sqrt{(E-T_{\text{core}})^2-m_\nu^2}/T_\nu}+1} \left(1+ \frac{v_{\text{esc}}^2}{v_{\text{avg}}^2}\right) \nonumber\\
    & \times \left(\frac{R_{\text{NS}}} {d_{\text{NS}}}\right)^2\sqrt{1-2\frac{G M_{\text{NS}}}{R_{\text{NS}}}}\,\Theta(E-m_\nu-T_{\text{core}})
\end{align}}
Note here we have assumed the resulting spectrum to be monochromatic at energies $T_{\text{core}}$. Although a realistic flux will have a finite width, but given the original monochromatic spectrum of the non-relativistic C$\nu$B neutrino flux as well as the Pauli-blocking of the final state neutron will lead to a finer width, therefore  assuming a monochromatic spectrum is a reasonable assumption for this boosted C$\nu$B flux.  We will look at two different applications of the differential flux derived above. Since, the flux decreases as the square of the distance to the candidate NS, we will first focus on the closest NS. The closest NS to the earth is NS from M7 i.e. RX J1856, which is about 123 pc away. The boosted flux from J1856 is shown in red in Fig.~\ref{fig:p3}. 

Next we consider the impact of a newly formed NS after the next galatic supernova, on the boosted flux arriving on Earth. As a typical candidate, we consider a Betelgeuse-type star undergoing SN explosion nearly 125-150 pc away. Although at the time of NS formation, the internal core temperature can reach upto 30-40 MeV but within few days it gradually cools down to internal temperatures around 100-200 keV. In the standard slow cooling scenario, these internal temperatures remains approximately steady for a initial period of about 2-3 years after NS formation~\cite{Sales:2020A}. The boosted flux from a potential nearby nascent NS is shown in orange in Fig.~\ref{fig:p3}. For comparison, we also include the expected boosted flux (shown in gray) from the NS1987A i.e. the compact remnant from SN1987A, currently believed to have internal temperature $\sim 30$ keV~\cite{Page:2020gsx}.  

We can see that all three different boosted C$\nu$B curves lie in the same energy range as solar pp and CNO neutrino flux. However, the boosted flux from J1856 and nearby future SN remnant is generically 4-8 orders of magnitude below the solar flux. These sources contribute a low-energy neutrino flux which is currently hard to detect. 
Since any newly formed NS (similarly NS1987A) will be in their neutrino cooling stage, the boosted C$\nu$B flux will be dwarfed by the neutrinos emitted from standard cooling. Due to the final state Pauli blocking of neutrons and protons, the spectrum for thermally emitted neutrinos is also peaked at core temperature. However the spectrum for thermal neutrinos is expected to be broader than for boosted C$\nu$B, due to the former being produced from three/four-body final state. It should also be pointed out that although the flux from J1856 is at low-energies and currently undetectable, it does not suffer from a background of thermally emitted neutrinos, as in the case of NS1987A or a newly-born NS.

The first ever real-time observation of pp-solar neutrinos was done by Borexino, reported in 2014~\cite{BOREXINO:2014pcl}. Borexino is a liquid scintillator detector with target mass $\sim 280$ ton, which detects solar neutrinos through neutrino-electron elastic scattering. The first measurement of CNO cycle solar neutrinos was also announced by Borexino in 2020~\cite{BOREXINO:2020aww}. The minimum energy threshold for the recoil electron in the Borexino detector is $\sim 50$ keV~\cite{BOREXINO:2018ohr}. The maximum recoil energy of the electron ($T_e$) for a given neutrino energy ($E_\nu$) is given by
\begin{equation}
    T_e^{\text{max}} = \frac{E_\nu}{1+\frac{m_e}{2E_\nu}}
\end{equation}
This implies $T_e^{\text{max}}\sim 50$ keV corresponds to $E_\nu\sim 140$ keV. Therefore, even in the case with large overdensities, the boosted flux from J1856 and NS1987A is currently unobservable. Although a future newly-born SN remnant might boost C$\nu$B above the current experimental threshold, given the relevant temperatures (100-200 keV), the electron recoil spectrum will only be modified at the lower energies which might not be an appreciable effect to constrain overdensities. However, this boosted flux might be observable in future with ultra-sensitive low-threshold detectors. Further discussion is outside the scope of current work and will be pursued in a future study. 


\section{Long-range forces}\label{sec:longrange}
In the previous section, we found that in the standard scenario the effect of NS cooling through C$\nu$B is negligible and the boosted C$\nu$B flux is currently beyond the detection capabilities of the terrestrial experiments. However, there could be more interesting effects at play in the presence of new physics beyond the Standard Model. In this section, we will focus our discussion on the effect of long-range forces. We will consider an ultra-light scalar $\phi$ that mediates a long-range force between nucleons and neutrinos. 

There can be two different cases of interest. In the first case,  presence of ultralight scalar coupling to neutrinos generates an attractive potential which can lead to the formation of stable C$\nu$B neutrino haloes which \GC{might be centered at the NS and} can have overdensities upto $\eta\sim 10^7$~\cite{Smirnov:2022sfo}. In this case, say for $m_\nu \sim 0.1$ eV, although the $L_\nu/L_\gamma$ would still be $\sim 10^{-3}$ but it could lead to a higher flux of boosted C$\nu$B on earth. \GC{ However, since it is not completely guaranteed for a neutrino halo to be centered around a NS, we do not delve further into the details of this case. }

In the second case, the attractive force between the C$\nu$B neutrinos and the neutrons in the NS can enhance the capture cross-section as well as boost the infalling neutrinos. We will focus on the latter case and shall also ensure that the boosting from the long-range force does not lead to the heating of the NS. Consider the following relevant Lagrangian terms for the interaction of nucleons $N$ and neutrinos $\nu$ with a ultralight scalar $\phi$, 
\begin{equation}
    \mathcal{L} \ \supset \ - y_{\nu} \overline{\nu}\, \phi\, \nu - y_\text{N} \bar{N}\phi N \, .
    \label{Dirac}
\end{equation}
where $y_\nu$ and $y_\text{N}$ are the coupling strengths of the neutrinos and nucleons to the scalar respectively. Therefore, the enhanced capture cross section is given by the expression 
\begin{equation}
    R_{\text{cap}} \simeq \left(1+\frac{2G M_{\text{NS}}}{R_{\text{NS}} v^2_{\text{avg}}} + \frac{2y_\nu\,y_\text{N}\,N_\text{N}}{4\pi m_\nu R_{\text{NS}} v^2_{\text{avg}}} \right)
\end{equation}
where $N_N$ is the number of nucleons in the NS. Note here we assume that the range of the ultralight scalar (i.e. $m_\phi^{-1}$) is longer than not only the typical NS size but also the enhanced capture radius $R_{cap}$. Similarly, the average boost velocity gets modified to
\begin{equation}
    v_{\text{boost}} \simeq \frac{1}{2}\sqrt{ \frac{G M_{\text{NS}}}{R_{\text{NS}}}+\frac{y_\nu\,y_\text{N}\,N_\text{N}}{4\pi m_\nu R_{\text{NS}}}}
\end{equation}
Given the current bounds on $y_\nu$ and $y_N$ coupled with large $N_\text{N}\sim 10^{57}$, we might expect the cooling rate to be substantially enhanced~\cite{Smirnov:2019cae,Babu:2019iml,Chauhan:2024qew}. However, there are few subtle points to be noted, which severely decreases the naive expectation of large enhancement. The presence of strong attractive potential due to $\phi$ not only increases the capture rate but also leads to an increase in the escape velocity at the NS surface. At first glance, we might be interested only in the energy exchange with the NS and even try to investigate the effect of trapped neutrino flux on the equilibrium conditions of the NS. However, due to the observation of thermal neutrino spectrum from SN1987A, we can constrain the escape velocity to be below unity. For example, let us consider the case of $m_\nu\sim 0.1$ eV. Requiring the escape velocity $v_{\text{esc}}\sim 2\sqrt{2}v_{\text{boost}} <1$, yields $y_\nu\,y_\text{N}<10^{-47}$. Using this coupling product to evaluate $L_\nu/L_\gamma$, we obtain $4.3 \times 10^{-11}$, which is only four times larger compared to the standard case.

\section{Conclusions}\label{sec:conclusion}
In this work, we have discussed about how large coherent enhancement can help C$\nu$B \textit{thermalize} with the NS.
The incoming low energy neutrino scatter off the high density nuclear matter. The medium sees the incoming neutrino has a long de Broglie wavelength. Therefore, incoming neutrino experiences large coherent volumes for its interaction/scattering rather than individual nucleons. Despite the Pauli blocking of the final state neutron, it is possible to exchange energy $\mathcal{O}(T)$ from the Fermi sphere. We have also discussed the theoretical framework to formally calculate the relevant cross-section, given the dynamic structure functions are available for such low-energy neutrino scattering in dense nuclear matter.   

Once the C$\nu$B gains energy, we predict a boosted flux on Earth and discuss about its observational prospect. We specifically discuss three different cases : J1856, NS1987A and future SN remnant. We find that this boosted flux might not be observable, unless in the presence of overdensities. We finally consider the case with long-range force mediated by an ultralight scalar. We establish that in case of neutrino haloes, the cooling might become appreciable but still sub-dominant along with a higher flux of boosted C$\nu$B neutrinos at Earth. While in the second case of neutrino-nucleon long-range interaction, we find a strong effect is constrained due to the escape velocity constraint from SN1987A. Although inclusion of long-range forces in the second case does not improve the bounds, we have outlined a new technique to constrain long-range forces using NSs.  

There are other several aspects to explore for future work including the future detection prospects for boosted C$\nu$B flux in ultra-sensitive low-threshold detectors, the diffuse boosted flux from all NSs, effect of redshift dependence, scattering off nuclear pasta phases which are known to have periodic cell structures as well as using NS cooling codes to study the effect of C$\nu$B cooling on the NS thermodynamic equilibrium in presence of overdensities and/or interactions through the magnetic moment portal~\cite{Akhmedov:2018wlf, Asteriadis:2022zmo}. 

\section*{Acknowledgements}\label{sec:acknowledgements}

We are very thankful to Gonzalo Herrera, Ian Shoemaker, Patrick Huber, Shunsaku Horiuchi, Andrew Gustafson, Diego Aristizabal Sierra, Ranjan Laha, Cecilia Lunardini and Jack Shergold for insightful comments and discussions. The work of GC is supported by the U.S. Department of Energy under the award number DE-SC0020250 and DE-SC0020262. GC also acknowledges the Center for Theoretical Underground Physics and Related Areas (CETUP* 2024) and the Institute for Underground Science at SURF for hospitality and for providing a stimulating environment, where this work was first presented (for slides, see \cite{GC:cetup2024}).
\newline
\newline
\noindent
\textbf{Note added:} We also would like to thank Bhupal Dev and Takuya Okawa for discussions and for informing us about their related work on C$\nu$B and neutron stars. We refer the interested readers to their work~\cite{Das:2024thc}. 

\bibliography{ref}

\providecommand{\href}[2]{#2}\begingroup\raggedright\begin{thebibliography}{10}

\bibitem{Weinberg:1962zza}
S.~Weinberg, \emph{{Universal Neutrino Degeneracy}}, \href{https://doi.org/10.1103/PhysRev.128.1457}{\emph{Phys. Rev.} {\bfseries 128} (1962) 1457}.

\bibitem{PTOLEMY:2018jst}
{\scshape PTOLEMY} collaboration, \emph{{PTOLEMY: A Proposal for Thermal Relic Detection of Massive Neutrinos and Directional Detection of MeV Dark Matter}},  \href{https://arxiv.org/abs/1808.01892}{{\ttfamily 1808.01892}}.

\bibitem{Betti:2018bjv}
M.~G. Betti et~al., \emph{{A design for an electromagnetic filter for precision energy measurements at the tritium endpoint}}, \href{https://doi.org/10.1016/j.ppnp.2019.02.004}{\emph{Prog. Part. Nucl. Phys.} {\bfseries 106} (2019) 120} [\href{https://arxiv.org/abs/1810.06703}{{\ttfamily 1810.06703}}].

\bibitem{PTOLEMY:2019hkd}
{\scshape PTOLEMY} collaboration, \emph{{Neutrino physics with the PTOLEMY project: active neutrino properties and the light sterile case}}, \href{https://doi.org/10.1088/1475-7516/2019/07/047}{\emph{JCAP} {\bfseries 07} (2019) 047} [\href{https://arxiv.org/abs/1902.05508}{{\ttfamily 1902.05508}}].

\bibitem{Stodolsky:1974aq}
L.~Stodolsky, \emph{{Speculations on Detection of the Neutrino Sea}}, \href{https://doi.org/10.1103/PhysRevLett.34.110}{\emph{Phys. Rev. Lett.} {\bfseries 34} (1975) 110}.

\bibitem{Duda:2001hd}
G.~Duda, G.~Gelmini and S.~Nussinov, \emph{{Expected signals in relic neutrino detectors}}, \href{https://doi.org/10.1103/PhysRevD.64.122001}{\emph{Phys. Rev. D} {\bfseries 64} (2001) 122001} [\href{https://arxiv.org/abs/hep-ph/0107027}{{\ttfamily hep-ph/0107027}}].

\bibitem{Yoshimura:2014hfa}
M.~Yoshimura, N.~Sasao and M.~Tanaka, \emph{{Experimental method of detecting relic neutrino by atomic de-excitation}}, \href{https://doi.org/10.1103/PhysRevD.91.063516}{\emph{Phys. Rev. D} {\bfseries 91} (2015) 063516} [\href{https://arxiv.org/abs/1409.3648}{{\ttfamily 1409.3648}}].

\bibitem{Eberle:2004ua}
B.~Eberle, A.~Ringwald, L.~Song and T.~J. Weiler, \emph{{Relic neutrino absorption spectroscopy}}, \href{https://doi.org/10.1103/PhysRevD.70.023007}{\emph{Phys. Rev. D} {\bfseries 70} (2004) 023007} [\href{https://arxiv.org/abs/hep-ph/0401203}{{\ttfamily hep-ph/0401203}}].

\bibitem{Bauer:2021uyj}
M.~Bauer and J.~D. Shergold, \emph{{Relic neutrinos at accelerator experiments}}, \href{https://doi.org/10.1103/PhysRevD.104.083039}{\emph{Phys. Rev. D} {\bfseries 104} (2021) 083039} [\href{https://arxiv.org/abs/2104.12784}{{\ttfamily 2104.12784}}].

\bibitem{Opher:1974drq}
R.~Opher, \emph{{Coherent scattering of cosmic neutrinos}}, {\emph{Astron. Astrophys.} {\bfseries 37} (1974) 135}.

\bibitem{Langacker:1982ih}
P.~Langacker, J.~P. Leveille and J.~Sheiman, \emph{{On the Detection of Cosmological Neutrinos by Coherent Scattering}}, \href{https://doi.org/10.1103/PhysRevD.27.1228}{\emph{Phys. Rev. D} {\bfseries 27} (1983) 1228}.

\bibitem{Shvartsman:1982sn}
B.~F. Shvartsman, V.~B. Braginsky, S.~S. Gershtein, Y.~B. Zeldovich and M.~Y. Khlopov, \emph{{POSSIBILITY OF DETECTING RELICT MASSIVE NEUTRINOS}}, {\emph{JETP Lett.} {\bfseries 36} (1982) 277}.

\bibitem{Shergold:2021evs}
J.~D. Shergold, \emph{{Updated detection prospects for relic neutrinos using coherent scattering}}, \href{https://doi.org/10.1088/1475-7516/2021/11/052}{\emph{JCAP} {\bfseries 11} (2021) 052} [\href{https://arxiv.org/abs/2109.07482}{{\ttfamily 2109.07482}}].

\bibitem{Domcke:2017aqj}
V.~Domcke and M.~Spinrath, \emph{{Detection prospects for the Cosmic Neutrino Background using laser interferometers}}, \href{https://doi.org/10.1088/1475-7516/2017/06/055}{\emph{JCAP} {\bfseries 06} (2017) 055} [\href{https://arxiv.org/abs/1703.08629}{{\ttfamily 1703.08629}}].

\bibitem{Hara:1980}
T.~Hara and H.~Sato, \emph{{Scattering of the Cosmic Neutrinos by High Energy Cosmic Rays}}, \href{https://doi.org/10.1143/PTP.64.1089}{\emph{Progress of Theoretical Physics} {\bfseries 64} (1980) 1089}.

\bibitem{Hara:1980mz}
T.~Hara and H.~Sato, \emph{{Elastic and Inelastic Scattering of the Relic Neutrinos by High-energy Cosmic Rays}}, \href{https://doi.org/10.1143/PTP.65.477}{\emph{Prog. Theor. Phys.} {\bfseries 65} (1981) 477}.

\bibitem{Ciscar-Monsalvatje:2024tvm}
M.~C\'\i{}scar-Monsalvatje, G.~Herrera and I.~M. Shoemaker, \emph{{Upper Limits on the Cosmic Neutrino Background from Cosmic Rays}},  \href{https://arxiv.org/abs/2402.00985}{{\ttfamily 2402.00985}}.

\bibitem{DeMarchi:2024zer}
A.~G. De~Marchi, A.~Granelli, J.~Nava and F.~Sala, \emph{{Relic Neutrino Background from Cosmic-Ray Reservoirs}},  \href{https://arxiv.org/abs/2405.04568}{{\ttfamily 2405.04568}}.

\bibitem{Herrera:2024upj}
G.~Herrera, S.~Horiuchi and X.~Qi, \emph{{Diffuse Boosted Cosmic Neutrino Background}},  \href{https://arxiv.org/abs/2405.14946}{{\ttfamily 2405.14946}}.

\bibitem{Bauer:2022lri}
M.~Bauer and J.~D. Shergold, \emph{{Limits on the cosmic neutrino background}}, \href{https://doi.org/10.1088/1475-7516/2023/01/003}{\emph{JCAP} {\bfseries 01} (2023) 003} [\href{https://arxiv.org/abs/2207.12413}{{\ttfamily 2207.12413}}].

\bibitem{Freedman:1973yd}
D.~Z. Freedman, \emph{{Coherent Neutrino Nucleus Scattering as a Probe of the Weak Neutral Current}}, \href{https://doi.org/10.1103/PhysRevD.9.1389}{\emph{Phys. Rev. D} {\bfseries 9} (1974) 1389}.

\bibitem{Wilson:1974zz}
J.~R. Wilson, \emph{{Coherent Neutrino Scattering and Stellar Collapse}}, \href{https://doi.org/10.1103/PhysRevLett.32.849}{\emph{Phys. Rev. Lett.} {\bfseries 32} (1974) 849}.

\bibitem{Lewis:1979mu}
R.~R. Lewis, \emph{{Coherent Detector for Low-energy Neutrinos}}, \href{https://doi.org/10.1103/PhysRevD.21.663}{\emph{Phys. Rev. D} {\bfseries 21} (1980) 663}.

\bibitem{Cabibbo:1982bb}
N.~Cabibbo and L.~Maiani, \emph{{The Vanishing of Order $G$ Mechanical Effects of Cosmic Massive Neutrinos on Bulk Matter}}, \href{https://doi.org/10.1016/0370-2693(82)90127-7}{\emph{Phys. Lett. B} {\bfseries 114} (1982) 115}.

\bibitem{Dixit:1983aj}
V.~V. Dixit and J.~Lodenquai, \emph{{ON NEUTRON STAR COOLING BY RELIC NEUTRINOS}}, \href{https://doi.org/10.1007/BF02789550}{\emph{Lett. Nuovo Cim.} {\bfseries 38} (1983) 174}.

\bibitem{Akhmedov:2018wlf}
E.~Akhmedov, G.~Arcadi, M.~Lindner and S.~Vogl, \emph{{Coherent scattering and macroscopic coherence: Implications for neutrino, dark matter and axion detection}}, \href{https://doi.org/10.1007/JHEP10(2018)045}{\emph{JHEP} {\bfseries 10} (2018) 045} [\href{https://arxiv.org/abs/1806.10962}{{\ttfamily 1806.10962}}].

\bibitem{COHERENT:2017ipa}
{\scshape COHERENT} collaboration, \emph{{Observation of Coherent Elastic Neutrino-Nucleus Scattering}}, \href{https://doi.org/10.1126/science.aao0990}{\emph{Science} {\bfseries 357} (2017) 1123} [\href{https://arxiv.org/abs/1708.01294}{{\ttfamily 1708.01294}}].

\bibitem{Bednyakov:2018mjd}
V.~A. Bednyakov and D.~V. Naumov, \emph{{Coherency and incoherency in neutrino-nucleus elastic and inelastic scattering}}, \href{https://doi.org/10.1103/PhysRevD.98.053004}{\emph{Phys. Rev. D} {\bfseries 98} (2018) 053004} [\href{https://arxiv.org/abs/1806.08768}{{\ttfamily 1806.08768}}].

\bibitem{Vitagliano:2019yzm}
E.~Vitagliano, I.~Tamborra and G.~Raffelt, \emph{{Grand Unified Neutrino Spectrum at Earth: Sources and Spectral Components}}, \href{https://doi.org/10.1103/RevModPhys.92.045006}{\emph{Rev. Mod. Phys.} {\bfseries 92} (2020) 45006} [\href{https://arxiv.org/abs/1910.11878}{{\ttfamily 1910.11878}}].

\bibitem{Trautner:2016ias}
A.~Trautner, \emph{{Massive Fermi Gas in the Expanding Universe}}, \href{https://doi.org/10.1088/1475-7516/2017/03/019}{\emph{JCAP} {\bfseries 03} (2017) 019} [\href{https://arxiv.org/abs/1612.07249}{{\ttfamily 1612.07249}}].

\bibitem{Sawyer:1975js}
R.~F. Sawyer, \emph{{Neutrino Opacity of Neutron Star Matter}}, \href{https://doi.org/10.1103/PhysRevD.11.2740}{\emph{Phys. Rev. D} {\bfseries 11} (1975) 2740}.

\bibitem{Iwamoto:1982zp}
N.~Iwamoto and C.~J. Pethick, \emph{{EFFECTS OF NUCLEON NUCLEON INTERACTIONS ON SCATTERING OF NEUTRINOS IN NEUTRON MATTER}}, \href{https://doi.org/10.1103/PhysRevD.25.313}{\emph{Phys. Rev. D} {\bfseries 25} (1982) 313}.

\bibitem{Reddy:1998hb}
S.~Reddy, M.~Prakash, J.~M. Lattimer and J.~A. Pons, \emph{{Effects of strong and electromagnetic correlations on neutrino interactions in dense matter}}, \href{https://doi.org/10.1103/PhysRevC.59.2888}{\emph{Phys. Rev. C} {\bfseries 59} (1999) 2888} [\href{https://arxiv.org/abs/astro-ph/9811294}{{\ttfamily astro-ph/9811294}}].

\bibitem{Horowitz:2006pj}
C.~J. Horowitz and A.~Schwenk, \emph{{The Neutrino response of low-density neutron matter from the virial expansion}}, \href{https://doi.org/10.1016/j.physletb.2006.09.042}{\emph{Phys. Lett. B} {\bfseries 642} (2006) 326} [\href{https://arxiv.org/abs/nucl-th/0605013}{{\ttfamily nucl-th/0605013}}].

\bibitem{Horowitz:2016gul}
C.~J. Horowitz, O.~L. Caballero, Z.~Lin, E.~O'Connor and A.~Schwenk, \emph{{Neutrino-nucleon scattering in supernova matter from the virial expansion}}, \href{https://doi.org/10.1103/PhysRevC.95.025801}{\emph{Phys. Rev. C} {\bfseries 95} (2017) 025801} [\href{https://arxiv.org/abs/1611.05140}{{\ttfamily 1611.05140}}].

\bibitem{Bedaque:2018wns}
P.~F. Bedaque, S.~Reddy, S.~Sen and N.~C. Warrington, \emph{{Neutrino-nucleon scattering in the neutrino-sphere}}, \href{https://doi.org/10.1103/PhysRevC.98.015802}{\emph{Phys. Rev. C} {\bfseries 98} (2018) 015802} [\href{https://arxiv.org/abs/1801.07077}{{\ttfamily 1801.07077}}].

\bibitem{Schuetrumpf:2020a}
B.~{Schuetrumpf}, G.~{Mart{\'\i}nez-Pinedo} and P.~G. {Reinhard}, \emph{{Survey of nuclear pasta in the intermediate-density regime: Structure functions for neutrino scattering}}, \href{https://doi.org/10.1103/PhysRevC.101.055804}{\emph{\prc} {\bfseries 101} (2020) 055804} [\href{https://arxiv.org/abs/1912.10510}{{\ttfamily 1912.10510}}].

\bibitem{Haberl:2006xe}
F.~Haberl, \emph{{The Magnificent Seven: Magnetic fields and surface temperature distributions}}, \href{https://doi.org/10.1007/s10509-007-9342-x}{\emph{Astrophysics} {\bfseries 308} (2007) 181} [\href{https://arxiv.org/abs/astro-ph/0609066}{{\ttfamily astro-ph/0609066}}].

\bibitem{Potekhin:2020ttj}
A.~Y. Potekhin, D.~A. Zyuzin, D.~G. Yakovlev, M.~V. Beznogov and Y.~A. Shibanov, \emph{{Thermal luminosities of cooling neutron stars}}, \href{https://doi.org/10.1093/mnras/staa1871}{\emph{Mon. Not. Roy. Astron. Soc.} {\bfseries 496} (2020) 5052} [\href{https://arxiv.org/abs/2006.15004}{{\ttfamily 2006.15004}}].

\bibitem{Buschmann:2019pfp}
M.~Buschmann, R.~T. Co, C.~Dessert and B.~R. Safdi, \emph{{Axion Emission Can Explain a New Hard X-Ray Excess from Nearby Isolated Neutron Stars}}, \href{https://doi.org/10.1103/PhysRevLett.126.021102}{\emph{Phys. Rev. Lett.} {\bfseries 126} (2021) 021102} [\href{https://arxiv.org/abs/1910.04164}{{\ttfamily 1910.04164}}].

\bibitem{Gould:1987ir}
A.~Gould, \emph{{Resonant Enhancements in WIMP Capture by the Earth}}, \href{https://doi.org/10.1086/165653}{\emph{Astrophys. J.} {\bfseries 321} (1987) 571}.

\bibitem{Goldman:1989nd}
I.~Goldman and S.~Nussinov, \emph{{Weakly Interacting Massive Particles and Neutron Stars}}, \href{https://doi.org/10.1103/PhysRevD.40.3221}{\emph{Phys. Rev. D} {\bfseries 40} (1989) 3221}.

\bibitem{Hansen:1997hb}
B.~M.~S. Hansen and E.~S. Phinney, \emph{{Stellar forensics I: Cooling curves}}, \href{https://doi.org/10.1046/j.1365-8711.1998.01232.x}{\emph{Mon. Not. Roy. Astron. Soc.} {\bfseries 294} (1998) 557} [\href{https://arxiv.org/abs/astro-ph/9708273}{{\ttfamily astro-ph/9708273}}].

\bibitem{Yakovlev:2004iq}
D.~G. Yakovlev and C.~J. Pethick, \emph{{Neutron star cooling}}, \href{https://doi.org/10.1146/annurev.astro.42.053102.134013}{\emph{Ann. Rev. Astron. Astrophys.} {\bfseries 42} (2004) 169} [\href{https://arxiv.org/abs/astro-ph/0402143}{{\ttfamily astro-ph/0402143}}].

\bibitem{Yakovlev:2004yr}
D.~G. Yakovlev, O.~Y. Gnedin, M.~E. Gusakov, A.~D. Kaminker, K.~P. Levenfish and A.~Y. Potekhin, \emph{{Neutron star cooling}}, \href{https://doi.org/10.1016/j.nuclphysa.2005.02.061}{\emph{Nucl. Phys. A} {\bfseries 752} (2005) 590} [\href{https://arxiv.org/abs/astro-ph/0409751}{{\ttfamily astro-ph/0409751}}].

\bibitem{Potekhin:MainReview}
A.~Y. {Potekhin}, J.~A. {Pons} and D.~{Page}, \emph{{Neutron Stars{\textemdash}Cooling and Transport}},  \href{https://arxiv.org/abs/1507.06186}{{\ttfamily 1507.06186}}.

\bibitem{Smirnov:2022sfo}
A.~Y. Smirnov and X.-J. Xu, \emph{{Neutrino bound states and bound systems}}, \href{https://doi.org/10.1007/JHEP08(2022)170}{\emph{JHEP} {\bfseries 08} (2022) 170} [\href{https://arxiv.org/abs/2201.00939}{{\ttfamily 2201.00939}}].

\bibitem{Batell:2024hzo}
B.~Batell and W.~Yin, \emph{{Cosmic Stability of Dark Matter from Pauli Blocking}},  \href{https://arxiv.org/abs/2406.17028}{{\ttfamily 2406.17028}}.

\bibitem{KATRIN:2022kkv}
{\scshape KATRIN} collaboration, \emph{{New Constraint on the Local Relic Neutrino Background Overdensity with the First KATRIN Data Runs}}, \href{https://doi.org/10.1103/PhysRevLett.129.011806}{\emph{Phys. Rev. Lett.} {\bfseries 129} (2022) 011806} [\href{https://arxiv.org/abs/2202.04587}{{\ttfamily 2202.04587}}].

\bibitem{Brdar:2022kpu}
V.~Brdar, P.~S.~B. Dev, R.~Plestid and A.~Soni, \emph{{A new probe of relic neutrino clustering using cosmogenic neutrinos}}, \href{https://doi.org/10.1016/j.physletb.2022.137358}{\emph{Phys. Lett. B} {\bfseries 833} (2022) 137358} [\href{https://arxiv.org/abs/2207.02860}{{\ttfamily 2207.02860}}].

\bibitem{Franklin:2024enc}
J.~Franklin, I.~Martinez-Soler, Y.~F. Perez-Gonzalez and J.~Turner, \emph{{Constraints on the Cosmic Neutrino Background from NGC 1068}},  \href{https://arxiv.org/abs/2404.02202}{{\ttfamily 2404.02202}}.

\bibitem{Sales:2020A}
T.~{Sales}, O.~{Louren{\c{c}}o}, M.~{Dutra} and R.~{Negreiros}, \emph{{Revisiting the thermal relaxation of neutron stars}},  \href{https://arxiv.org/abs/2004.05019}{{\ttfamily 2004.05019}}.

\bibitem{Page:2020gsx}
D.~Page, M.~V. Beznogov, I.~Garibay, J.~M. Lattimer, M.~Prakash and H.-T. Janka, \emph{{NS 1987A in SN 1987A}}, \href{https://doi.org/10.3847/1538-4357/ab93c2}{\emph{Astrophys. J.} {\bfseries 898} (2020) 125} [\href{https://arxiv.org/abs/2004.06078}{{\ttfamily 2004.06078}}].

\bibitem{BOREXINO:2014pcl}
{\scshape BOREXINO} collaboration, \emph{{Neutrinos from the primary proton\textendash{}proton fusion process in the Sun}}, \href{https://doi.org/10.1038/nature13702}{\emph{Nature} {\bfseries 512} (2014) 383}.

\bibitem{BOREXINO:2020aww}
{\scshape BOREXINO} collaboration, \emph{{Experimental evidence of neutrinos produced in the CNO fusion cycle in the Sun}}, \href{https://doi.org/10.1038/s41586-020-2934-0}{\emph{Nature} {\bfseries 587} (2020) 577} [\href{https://arxiv.org/abs/2006.15115}{{\ttfamily 2006.15115}}].

\bibitem{BOREXINO:2018ohr}
{\scshape BOREXINO} collaboration, \emph{{Comprehensive measurement of $pp$-chain solar neutrinos}}, \href{https://doi.org/10.1038/s41586-018-0624-y}{\emph{Nature} {\bfseries 562} (2018) 505}.

\bibitem{Smirnov:2019cae}
A.~Y. Smirnov and X.-J. Xu, \emph{{Wolfenstein potentials for neutrinos induced by ultra-light mediators}}, \href{https://doi.org/10.1007/JHEP12(2019)046}{\emph{JHEP} {\bfseries 12} (2019) 046} [\href{https://arxiv.org/abs/1909.07505}{{\ttfamily 1909.07505}}].

\bibitem{Babu:2019iml}
K.~S. Babu, G.~Chauhan and P.~S. Bhupal~Dev, \emph{{Neutrino nonstandard interactions via light scalars in the Earth, Sun, supernovae, and the early Universe}}, \href{https://doi.org/10.1103/PhysRevD.101.095029}{\emph{Phys. Rev. D} {\bfseries 101} (2020) 095029} [\href{https://arxiv.org/abs/1912.13488}{{\ttfamily 1912.13488}}].

\bibitem{Chauhan:2024qew}
G.~Chauhan and X.-J. Xu, \emph{{Impact of the cosmic neutrino background on long-range force searches}}, \href{https://doi.org/10.1007/JHEP07(2024)255}{\emph{JHEP} {\bfseries 07} (2024) 255} [\href{https://arxiv.org/abs/2403.09783}{{\ttfamily 2403.09783}}].

\bibitem{Asteriadis:2022zmo}
K.~Asteriadis, A.~Q. Trivi\~no and M.~Spinrath, \emph{{Bremsstrahlung from neutrino scattering via magnetic dipole moments}}, \href{https://doi.org/10.1142/S0217751X23501397}{\emph{Int. J. Mod. Phys. A} {\bfseries 38} (2023) 2350139} [\href{https://arxiv.org/abs/2208.01207}{{\ttfamily 2208.01207}}].

\bibitem{GC:cetup2024}
G.~Chauhan, ``\textit{Neutron Stars as probe of Cosmic Neutrino Background}.'' \href{https://indico.sanfordlab.org/event/69/contributions/1468/attachments/889/2202/CosmicNeutrinoBackground_Chauhan.pdf}{CETUP* slides}, July 2024.

\bibitem{Das:2024thc}
S.~Das, P.~S.~B. Dev, T.~Okawa and A.~Soni, \emph{{Old neutron stars as a new probe of relic neutrinos and sterile neutrino dark matter}}, \href{https://doi.org/10.1103/PhysRevD.111.055035}{\emph{Phys. Rev. D} {\bfseries 111} (2025) 055035} [\href{https://arxiv.org/abs/2408.01484}{{\ttfamily 2408.01484}}].

\end{thebibliography}\endgroup

\end{document}